# Control of Photon Dynamics in Non-Euclidean Polygonal Microcavities by Joint Geometric Curvatures

Yechun Ding,[1] Yongsheng Wang,[1,2] Peng Li,[3] Yaxin Guo[1], Yanpeng Zhang,[1] Feng Yun,[1,4] and Feng Li[1,4,*]

[1] Key Laboratory for Physical Electronics and Devices of the Ministry of Education & Shaanxi Key Lab of Information Photonic Technique, School of Electronic Science and Engineering, Faculty of Electronic and Information Engineering, Xi'an Jiaotong University, Xi'an 710049, China;

[2] Department of Electrical and Photonics Engineering, Technical University of Denmark, Kgs. Lyngby, 2800, Denmark;

[3] Institude of Regenerative and Reconstructive Medicine, Med-X Institute, The First Affiliated Hospital of Xi'an Jiaotong University, Xi'an 710061, China;

[4] Solid-State Lighting Engineering Research Center, Xi'an Jiaotong University, Xi'an 710049, China

**ABSTRACT**. Non-Euclidean geometry has recently emerged as a powerful tool, offering new insights and applications in optical microcavities supporting Whispering Gallery Modes (WGMs). In this study, we extend the concept of polygonal microcavities to non-Euclidean spaces by developing a unified model that incorporates a joint geometric parameter of curvatures. This system uncovers a range of unexplored phenomena, mechanisms, and concepts that are unique to curved spaces. Notably, we observe dissipative states characterized by hyperbolic fixed points (HFPs) that appear exclusively in non-Euclidean scenarios, leading to the formation of phase diagrams within the parametric space of curvatures. Our results reveal phase transitions across geometric boundaries, marked by abrupt changes in the cavity quality factor. These transitions are strongly influenced by the wavelike nature of photon trajectories, offering intriguing insights into quantum chaos within curved spaces. Additionally, we discover that cavities with geodesic side lines exhibit a remarkable symmetry-driven avoidance of such phase transitions, highlighting the profound connection between physical dynamics and spatial geometry. Our findings establish a promising platform for optical simulations of non-Euclidean quantum chaos and open up potential applications in on-chip photonic devices.

*Introduction*—Optical microcavities supporting Whispering-gallery modes (WGMs) have garnered significant attention as essential components in integrated photonics, owing to their applications in on-chip coherent and quantum light sources [1-5], filters and isolators [6-8], nonlinear optical platforms [9-11], and high-performance sensors [23,24]. The primary advantage of WGM microcavities lies in their high quality factor (Q-factor), which is facilitated by total internal reflection (TIR) at the circular cavity boundaries. Additionally, unidirectional emission is achieved by introducing slight deformations to the cavity boundaries, thereby creating chaotic tunneling channels [12].The utilization of chaotic photon dynamics, resulting from broken rotational symmetry, has led to the observation of various exotic phenomena and applications, such as broadband momentum coupling [13,14], broadband frequency combs [15,16], and phase-space tailoring [17]. Recently, the investigation of WGM photon dynamics has extended to non-Euclidean spaces by defining microcavities on curved surfaces, including Möbius strip microlasers [18] and microcavities on surfaces of revolution [19]. This extension offers an additional degree of freedom for light manipulation, revealing emergent physics in higher-dimensional spaces.

An important question in the study of non-Euclidean metrics is how the curvature of space influences photon dynamics within cavities that have been thoroughly investigated in flat space. This line of inquiry was initially explored in a Face cavity defined on a spherical surface by continuously varying the surface curvature [20]. The study demonstrated that positive space curvature significantly reduces dissipation in low-periodic modes, leading to enhanced mode coupling and the generation of non-Hermitian exceptional points (EPs) [20]. Furthermore, it was found that regular and chaotic modes respond differently to changes in space curvature, resulting in the formation of localized hybrid wavepackets that are not possible in flat space [21].

In addition to the near-circular cavities discussed previously, polygonal microcavities have also been extensively studied for both fundamental physics and practical applications [25-28]. The appeal of polygonal cavities lies in the intrinsic relationship between their geometry and the associated physical phenomena. The investigation of geometry-dependent mechanisms has spanned several decades, and it was only recently demonstrated that universal criteria exist to determine whether an arbitrarily shaped polygon can support Whispering-gallery modes (WGMs) [22]. The polygonal shape makes such cavities particularly promising for directional laser emission [25,26] and enables the bottom-up fabrication of micro and nanolasers [29-35].

*Contact author: felix831204@xjtu.edu.cn

To address the disadvantage of relatively low Q-factors arising from pseudo-integrable photon dynamics, designs incorporating circular-sided polygons have been proposed to improve the stability of optical modes, albeit at the cost of introducing chaotic dynamics [36]. In this context, it is crucial to understand whether the principles governing polygonal microcavities, featuring both straight and curved sides, can be extended to curved space. Can the concept of a polygonal microcavity be applied to non-Euclidean spaces? How would the photon dynamics in such cavities respond to the curvature of space? Are there fundamental differences between cavities with geodesic and non-geodesic sides? Finally, can the dynamics of Euclidean cavities be regarded as a special case within the broader framework of non-Euclidean systems?

In this letter, we present a universal model for non-Euclidean polygonal microcavities, featuring both straight and circular sides, which act as optical simulators for classical (ray) and quantum (wave) chaos. This model is applicable to investigations involving both ray and wave dynamics [21]. We demonstrate that the regular motion of cavity photons strongly depends on the combined geometry of space and boundary curvatures, resulting in both stable and unstable fixed points in phase space. Distinct phase transitions occur between different geometric regions, characterized by a sharp variation in the cavity quality factor (Q-factor), which stems from the reduced stability of chaotic dynamics. Remarkably, cavities with geodesic sides exhibit a unique feature: symmetry-driven avoidance of phase transitions, highlighting the fundamental difference between geodesic and non-geodesic cavities. Furthermore, we observe a striking difference between classical and quantum chaos near the phase transition point. While strong variations in dissipation are seen at short wavelengths, this effect is completely mitigated when the dynamics exhibit a sufficiently wavelike behavior. These findings extend the concept of polygonal microcavities to non-Euclidean spaces, offering valuable insights into the fundamental physics of higher-dimensional systems. This work holds potential for the design of future photonic devices that exploit chaotic dynamics with coupled degrees of freedom.

*Theoretical model*—We begin by considering a regular polygonal cavity defined on a hemispherical surface, as depicted in Fig. 1(a). Let $R_s$ denote the radius of curvature of the spherical surface, with $O$ representing the center of the sphere. The center of the polygonal surface is denoted as $O''$, and the line connecting $O$ and $O''$ is taken as the $z$-axis. $R_e$ represents the distance between the vertex (denoted as $A_i$, $i=1, 2,...N$ in which $N$ is the number of sides) of the polygon and the $z$-axis, which intersects with the $z$-axis at the point O'. To maintain a constant cavity size, we set $R_e = 2$ μm. The effective curvature of space, denoted by $\eta$, which is defined by $\sin(\eta \cdot \pi/2) = R_e/R_s$, is inversely proportional to the radius of curvature $R_s$ and is confined within the range $0 < \eta < 1$. The key challenge in this setup is the definition of curvatures of the sides, as all lines on the spherical surface are curved from a Euclidean perspective. This issue is addressed by introducing a variable $z_m$, which represents the length of $OC$, where $C$ is a point on the segment $OO'$. The point $C$ and two adjacent vertices form a plane $CA_iA_{i+1}$, which intersects the spherical surface. The curve of intersection defines the side connecting the $i_{th}$ and $(i+1)_{th}$ vertices, as illustrated by the dotted lines in Fig. 1(a). We then define the effective curvature of the side $\zeta = z_m/L$, where L is the length of the segment OO'. The variable $\zeta$ is constrained within the range $0 \leq \zeta \leq 1$, with $0 \leq z_m \leq L$. Notably, when $\zeta = 0$ (i.e., C coincides with O), the sides of the polygon are great circles, i.e., geodesic lines on the spherical surface. In this case, the sides of the polygon remain geodesic for varying values of $\eta$ and become straight lines when $\eta = 0$ (which corresponds to flat space), resulting in a Euclidean polygon with straight sides. Conversely, when $\zeta = 1$ (i.e., C coincides with $O'$), the curvature of the sides is sufficiently large such that the polygon becomes a circle on the spherical surface. The boundary remains circular for varying values of $\eta$, and the enclosed shape reduces to a Euclidean circle when $\eta = 0$. For intermediate values of ($0 < \zeta < 1$), the polygon has non-geodesic sides, which reduce to Euclidean polygons with curved sides (as previously discussed in Ref. [36]) when $\eta = 0$.

In this framework, the concept of a polygonal cavity is extended from Euclidean to non-Euclidean spaces, with the geometry fully described by the parameters $\eta$ and $\zeta$, collectively denoted as the joint geometric parameter of curvatures $G = (\eta, \zeta)$. As such, the terms

*Contact author: felix831204@xjtu.edu.cn

"straight" (resp. "curved") should be replaced by "geodesic" (resp. "non-geodesic"), and we define a polygon as geodesic (resp. non-geodesic) if its sides are (resp. are not) geodesic lines. As $\eta$ (resp. $\zeta$) increases, the integrability of the system (excluding polygons with three or four sides) increases, ultimately reaching a value of 1, at which the cavity becomes the surface of a hemisphere (resp. a spherical cap).

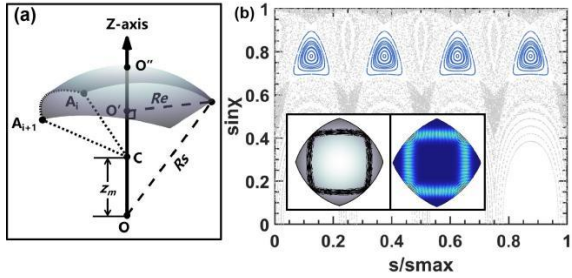

FIG. 1. Theoretical model and photon dynamics of the non-Euclidean polygonal microcavity. (a) Schematic of the microcavity. $O$ and $O''$ denotes the center of the sphere and the polygon, respectively. $A_i$ and $A_{i+1}$ denote the $i_{th}$ and $(i+1)_{th}$ vertices of the polygonal cavity, respectively. $O'$ represents the perpendicular foot of the vertex on the z-axis, with $C$ being a point along $OO'$. (b) Poincaré surface of section (PSOS) for a non-Euclidean polygonal microcavity with $\eta=0.5$ and $\zeta=0.3$. The blue points correspond to 4-periodic motion, while the gray points indicate chaotic trajectories. Insets show light trajectories (left) and optical field (right) for the 4-periodic islands.

*Results and discussion*—To begin the analysis, we consider quadrilateral cavities for simplicity, with the understanding that the results can be extended to polygonal cavities with more sides. Specifically, the sketch in Fig. 1(a) illustrates a square cavity. In typical calculations, the ray trajectories and wave distribution for an eigenstate are shown in the insets of Fig. 1(b) for $\eta = 0.5$ and $\zeta = 0.3$. The ray dynamics are calculated using the same algorithm as in our previous works [20, 21], while the wave dynamics are simulated using COMSOL Multiphysics (Version 5.6).

The calculated Poincaré surface of section (PSOS) is shown in Fig. 1(b), which reveals four-period islands (indicated in blue) surrounded by a chaotic sea. To study the impact of varying the curvature parameters $\eta$ and $\zeta$, we gradually adjust these parameters and monitor the changes in both the shapes and positions of the islands. As illustrated in Fig. 2(a), the shape of the islands transitions with varying $\zeta$ through a series of oval, inverted triangle, triangle, square, diamond, and multi-sided (with more than four sides) polygons, eventually returning to an oval shape. Throughout this process, the positions of the islands remain at the same height in the PSOS. We refer to these different island shapes as polygonal fixed points (PFPs) in the PSOS. It is important to note that the ovals observed at $\zeta = 0$ and $\zeta = 1$ represent distinctly different features, corresponding to polygonal shapes with two sides ($\zeta = 0$) and an infinite number of sides ($\zeta = 1$), respectively. Additionally, as $\eta$ increases, a similar variation in the shape of the islands occurs, similar to the effect of increasing $\zeta$. However, the positions of the islands shift upward, and the shapes become more compressed at higher values of $\eta$, as shown in Fig. 2(b). This behavior is a typical effect of the increasing curvature of the space, as already discussed in our previous work [20]. It is noteworthy that transient states of hyperbolic fixed points (HFPs) appear at the transition between different shapes of polygonal fixed points (PFPs). Specifically, a set of six-branch (resp. eight-branch) HFPs occurs at the transition between inverted triangles (resp. squares) and triangles (resp. diamonds). Detailed information on the HFPs is provided in the Supplementary Information. It is evident that the symmetry of the HFPs, characterized by the number of their branches, aligns perfectly with the connecting PFPs. In contrast to PFPs, HFPs represent unstable states that lead to the divergence of photon dynamics into the chaotic sea, thereby reducing the Q-factor.

We have extended our analysis to polygons with more than four sides, observing the same evolution of island shapes as in quadrilateral cavities (see Supplementary Information). This suggests that the shape transition is an inherent property of polygonal cavities, independent of the number of sides. Interestingly, the exact geometric parameters $G = (\eta, \zeta)$ at which the HFPs occur can be precisely determined by analytically calculating the eigenvalues of the Jacobian matrix, which is generally expressed as:

$$J(x_n) = \begin{bmatrix} \dfrac{\partial p_{n+1}}{\partial p_n} & \dfrac{\partial p_{n+1}}{\partial \theta_n} \\ \dfrac{\partial \theta_{n+1}}{\partial p_n} & \dfrac{\partial \theta_{n+1}}{\partial \theta_n} \end{bmatrix} \quad (1)$$

where $x_n=(\theta_n, p_n)$ represents the coordinates of one of the four fixed points in the PSOS at the $n_{th}$ reflection, and $x_{n+1}=(\theta_{n+1}, p_{n+1})$ denotes the subsequent mapped location at the $(n+1)_{th}$ reflection. In quadrilateral geometry, the relation between $x_n$ and $x_{n+1}$ is calculated and incorporated into Eq. (1), from which we obtain the eigenvalues of the Jacobian matrix. If the eigenvalues are real, the fixed points are unstable, indicating the presence of hyperbolic fixed points (HFPs) [37]. By recording the joint geometric parameters $G = (\eta, \zeta)$ at which the fixed point transitions to a hyperbolic state, we construct a two-dimensional (2D) phase diagram, as shown in Fig. 2(c). The red and blue dotted lines, corresponding to the geometric positions of the six-branch and eight-branch HFPs, respectively, divide the phase diagram into three regions based on the number of sides of the island polygons: two to three sides in Region I, three to four sides in Region II, and four to infinite sides in Region III. Any phase transition across the HFP lines induces a drop and subsequent recovery of system stability. It is important to note that the transition is not abrupt; instead, it is characterized by a gradual evolution of the size and shape of the islands as the parameters $\eta$ and $\zeta$ change. In fact, the HFPs form "bands" rather than distinct lines, with the lines in the phase diagram corresponding only to the exact solutions provided by the Jacobian matrix. In the absence of HFPs, the system's stability (or integrability) generally increases from the bottom-left to the top-right as the positive curvature increases. We computed the phase diagrams for polygonal cavities with more sides and found that all exhibit similar features. The phase diagram for the hexagon is shown in Fig. 2(d), while those for the triangle, pentagon, and octagon are presented in the Supplementary Information.

*Contact author: felix831204@xjtu.edu.cn

Significant features can be identified by further examining the phase diagrams of various polygonal cavities. First, we observe that in Euclidean space (i.e., when $\eta = 0$), no hyperbolic fixed point (HFP) exists for $\zeta < 1$. This suggests that the presence of HFPs in polygonal cavities, along with the associated phase transitions, is a distinctive feature arising from non-Euclidean geometry. Additionally, we find that the axis defined by $\zeta = 0$ (i.e., the $\eta$-axis) asymptotically approaches, but never crosses, the six-branch (resp. eight-branch) HFP line as $\eta$ approaches 1 for the hexagonal (resp. quadrilateral) cavity. The HFP is only reached when $\eta = 1$. It is noteworthy that such a "just-avoided-crossing" (JAC) behavior occurs exclusively when the geometry of the cavity and the HFPs share the same symmetry—namely, hexagon versus six-branch and quadrilateral versus eight-branch. This phenomenon does not occur in polygons with different symmetries, as illustrated in the phase diagrams shown in the Supplementary Information. Therefore, the JAC behavior is a key feature of geodesic polygonal cavities ($\zeta = 0$), reflecting a profound connection between geometric symmetry and physical motion, even in chaotic regimes. This behavior offers a robust criterion for distinguishing geodesic cavities from non-geodesic ones, a distinction that would not have been apparent without extending the study of polygonal cavities to non-Euclidean space.

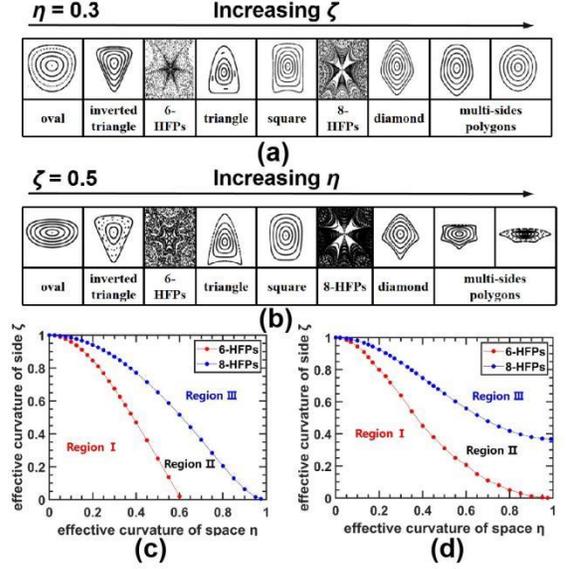

FIG. 2. Island morphology of the non-Euclidean polygonal microcavity under the variation of geometric curvatures. (a), (b) The island shape varies with different geometric parameters while keeping $\eta$=0.3 (a) and $\zeta$=0.5 (b) as constants, respectively. (c), (d) Phase diagrams for quadrilateral (a) and hexagonal (b) cavities. The 6-HFPs (red dashed lines) and 8-HFPs (blue dashed lines) partition the phase diagram into three regions: Region I, II, and III.

We now consider practical optical devices analyzed using wave optics, which inherently involve wave-like uncertainty and coherence. Consequently, the system can be viewed as an optical simulator of quantum chaos, as discussed in previous works [21]. The simulation is performed using COMSOL Multiphysics (Version 5.6), as in our earlier studies [20,21]. In the present work, however, we employ the principle of transformation optics to establish an equivalent 2D cavity with a gradient refractive index distribution, as depicted in Fig. 3(a). This approach enables higher precision through the use of a finer simulation mesh density. The validity of this technique has been rigorously demonstrated in numerous 3D objects, including curved cavities [39-42]. The corresponding refractive index distribution for the spherical surface is provided [38]:

$$n_{2D}(r) = n_{3D} \cdot \frac{2 \cdot R_0/C}{1 + (r/C)^2} \qquad (2)$$

where $R_0$ is spherical radius and $C$ is an integral constant which is set as

*Contact author: felix831204@xjtu.edu.cn

$$C = \frac{Re}{\tan(\eta \cdot \pi/4)} \qquad (3)$$

and $r$ is the 2D radial coordinate given by $r = C \cdot \tan(\theta/2)$. Here, $\theta$ represents the elevation angle of the sphere. The value of $C$ is chosen to ensure that the refractive index reaches its minimum $n_{2D\,min} = n_{3D}$, while the background refractive index is set to that of air. Additional mathematical details are provided in the Supplementary Information. The spatial distribution of the refractive index for a quadrilateral cavity with $\eta = 0.3$ and $\zeta = 0.5$ is shown in Fig. 3(b).

We calculate the Q-factors of the cavity around the high-frequency points (HFPs). First, we select the light frequency within a small range around 600 THz (with azimuthal number $m = 88$) and compute the Q-factor as a function of $\zeta$ for three values of $\eta$: $\eta = 0.3, 0.5,$ and 0.7. The results are shown in Fig. 3(c). Distinct valleys, characterized by orders-of-magnitude drops in the Q-factor (with the valley minima circled in the graph), appear, which directly correspond to the ($\eta, \zeta$) parameters of the HFPs in the phase diagram of Fig. 2(c). These valleys are fully expected due to the unstable nature of the HFPs, and they robustly demonstrate that chaotic ray dynamics accurately predict the performance of optical devices. In contrast, when the valleys are ignored, the Q-factor increases by several orders of magnitude as $\zeta$ and $\eta$ increase, which is consistent with previous studies [20,36]. Specifically, the positive curvature of the boundary and space generally enhances the stability of light trajectories, while positive space curvature shifts the islands upward in the phase-space trajectory (PSOS), both contributing to the increase in the Q-factor. When the light behaves more like a wave, quantum (wave) dynamics is expected to differ from classical (ray) dynamics. This is confirmed by examining the resonant modes at different optical frequencies, as shown in Fig. 3(d), where four frequency ranges (approximately 1200, 600, 300, and 200 THz, with $m = 188, 88, 48,$ and 36, respectively) are considered for $\eta = 0.3$. It is observed that the valleys in the Q-factor at the HFPs become shallower as the optical wavelength increases (or equivalently, the light frequency decreases), and are nearly flattened when the frequency reduces to approximately 200 THz. In terms of dissipation, the difference between stable photonic fixed points (PFPs) and unstable HFPs becomes negligible due to the wavelike uncertainty, allowing the photons to bypass the anomalous states in the classical (ray) phase diagram by operating in the quantum (wavelike) regime.

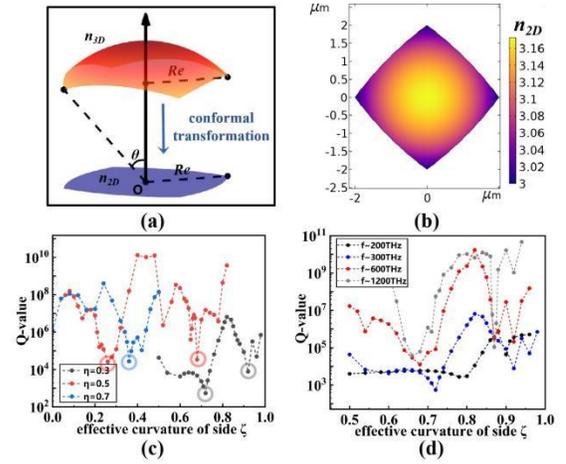

FIG. 3. Analysis of 4-periodic motions using wave dynamics. (a) Schematic of the transformation optics from a 3D cavity to a 2D cavity. $n_{3D}$: Uniform refractive index of the hemispherical polygonal microcavity; $n_{2D}$: Graded refractive index of the planar polygonal microcavity. (b) Schematic diagram showing the refractive index gradient in the 2D polygonal microcavity. (c) Q-factors as a function of $\zeta$ for $\eta$=0.3(black), 0.5 (red), and 0.7 (blue) within a narrow frequency range around 600 THz (azimuthal number m=88). The valley minima for $\eta$=0.3, $\eta$=0.5, and $\eta$=0.7 are highlighted by black, red, and blue circles, respectively. (d) Q-factors as a function of $\zeta$ for four different optical frequency ranges (around 1200, 600, 300, and 200 THz, with $m$=188,88,48, and 36), while keeping $\eta$=0.3.

In order to fully understand the mechanism of wavelike dissipation, we compare the Husimi projections [43,44] of the optical modes between the six-branch HFPs and the oval-like PFPs, as shown in Fig. 4. The insets display the corresponding real-space field distributions. The instability of the HFPs is clearly revealed by the leakage of the Husimi wavepacket along the directions of the branches (indicated by dashed arrows in Fig. 4(a) to 4(c)), forming fixed dissipation channels. At short wavelengths ($m$ = 188), although the Husimi wavepacket is predominantly localized at the island centers, non-negligible leakage through the dissipation channels reduces the Q-factor, resulting in directional emission at specific points along the cavity periphery when the system operates below the total internal reflection (TIR) line (indicated by horizontal dashed lines). In contrast, the wavepacket of the PFP is perfectly confined to the centers of the islands, benefiting from the high stability of the classical dynamics. As the light transitions to more wavelike behavior, the situation evolves. At $m$ = 88, the wavepacket spreads from the island center into the chaotic sea for both the HFP and PFP. However, this spreading remains relatively trivial compared to the dissipation channel of the PFP, which allows the PFP to maintain its superior Q-factor. As the wavelength increases further to $m = 36$, the Husimi wavepackets spread more significantly, leading to substantial dissipation into the chaotic sea. Even for the PFP, which is classically stable, the wavepacket begins to spread outside the island regions. This wavelike dissipation mechanism becomes dominant, surpassing the leakage from the classical HFP dissipation channels. This results in the characteristic bypassing of classical (ray-like) anomalous states, which can be interpreted as an optical analog of quantum (wave-like) properties.

*Contact author: felix831204@xjtu.edu.cn

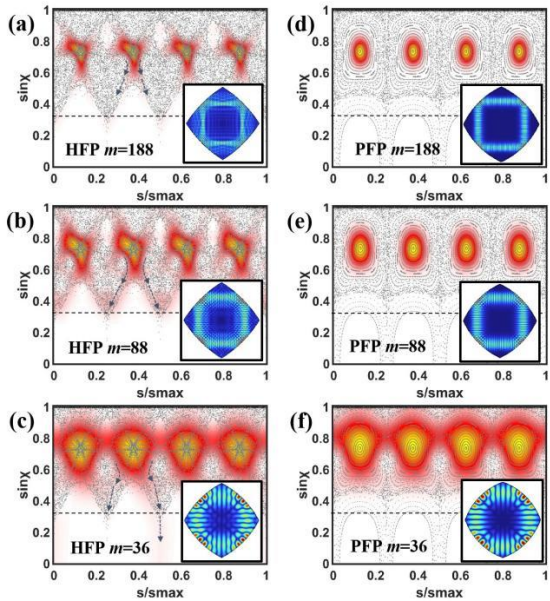

FIG. 4. Husimi projections (red shading) of the 4-period island mode for various values of $\zeta$ and azimuthal numbers $m$. (a) (d) $m$=188 HFP, (b) (e) $m$=88 HFP, (c) (f) $m$=36 for HFP and PFP, respectively. The spatial field distributions of each mode are shown as insets in each graph.

*Conclusion*—In conclusion, we extend the concept of polygonal microcavities to non-Euclidean spaces and uncover a series of consistent physical principles, including the HFP-PFP phase transition, the JAC criterion for geodesic cavities, and loss modulation via wave chaos, among others. Utilizing the framework of transformation optics, we demonstrate that the effect of spatial curvature can be realized through the inhomogeneity of the effective refractive index, which can be readily implemented using metasurfaces for on-chip photonic applications. This paves the way for non-Euclidean cavities to serve as a promising platform for the development of next-generation microlasers and microwave resonators. Specifically, microlasers can be engineered to exhibit local and directional emission by exploiting the dissipation channels of HFPs. Furthermore, ultra-high-Q polygonal microcavities can be designed by carefully tuning both the boundary and the space curvatures, taking advantage of the multiple degrees of freedom and the precise avoidance of lossy geometries of HFPs. Fundamentally, our system is shown to function as an ideal optical simulator for quantum chaos, offering valuable insights for high-dimensional geometry, dissipative topological physics, and cosmological studies.

### ACKNOWLEDGMENTS
This research was supported by National Key R&D Program of China (Grant No. 2023YFA1407100), the National Natural Science Foundation of China (NSFC) (Grant Nos. 12474392 and 12074303), and the Postdoctoral Fellowship Program of China Postdoctoral Science Foundation (Grant No. GZC20241349).

*Contact author: felix831204@xjtu.edu.cn

*Contact author: felix831204@xjtu.edu.cn

*Contact author: felix831204@xjtu.edu.cn

# Supplementary Information for "Control of Photon Dynamics in Non-Euclidean Polygonal Microcavities by Joint Geometric Curvatures"

Yechun Ding[1], Yongsheng Wang[1,2], Peng Li[3], Yaxin Guo[1], Yanpeng Zhang[1], Feng Yun[1,4] and Feng Li[1,4*]

[1] *Key Laboratory for Physical Electronics and Devices of the Ministry of Education & Shaanxi Key Lab of Information Photonic Technique, School of Electronic Science and Engineering, Faculty of Electronic and Information Engineering, Xi'an Jiaotong University, Xi'an 710049, China;*

[2] *Department of Electrical and Photonics Engineering, Technical University of Denmark, Kgs. Lyngby, 2800, Denmark;*

[3] *Institude of Regenerative and Reconstructive Medicine, Med-X Institute, The First Affiliated Hospital of Xi'an Jiaotong University, Xi'an 710061, China;*

[4] *Solid-State Lighting Engineering Research Center, Xi'an Jiaotong University, Xi'an 710049, China*

1. **Full PSOSs of quadrilateral and hexagonal cavities.**

We present the evolution of the shapes of the polygonal islands with the joint geometry curvatures, as shown in Figs. S1, S2 for quadrilateral and Figs. S3, S4 for hexagonal cavities. It's obvious to see that the shape of the islands transitions with varying $\zeta$ ($\eta$) through a series of oval, inverted triangle, 6-branch hyperbolic, triangle, square, 8-branch hyperbolic, diamond, and multi-sided (with more than four sides) polygons, eventually returning to an oval shape.

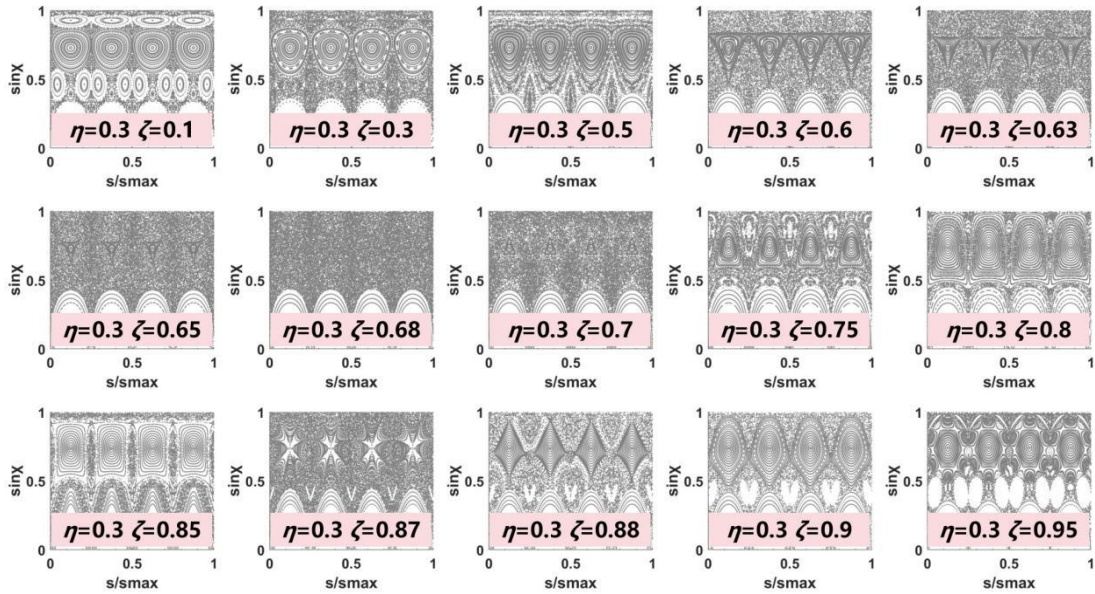

Fig. S1. The full PSOSs of quadrilateral cavities when keeping $\eta$ = 0.3 and varying $\zeta$ as 0.1, 0.3, 0.5, 0.6, 0.63, 0.65, 0.68, 0.7, 0.75, 0.8, 0.85, 0.87, 0.88, 0.9 and 0.95.

*Contact author: felix831204@xjtu.edu.cn

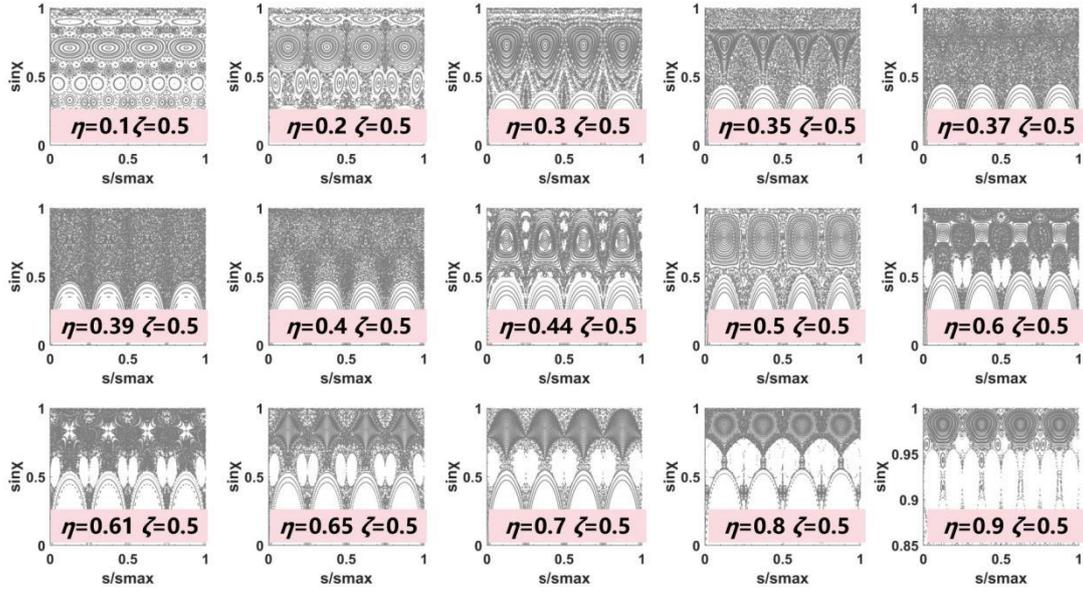

Fig. S2. The full PSOSs of quadrilateral cavities when keeping $\zeta = 0.3$ and varying $\eta$ as 0.1, 0.2, 0.3, 0.35, 0.37, 0.39, 0.4, 0.44, 0.5, 0.6, 0.61, 0.65, 0.7, 0.8 and 0.9.

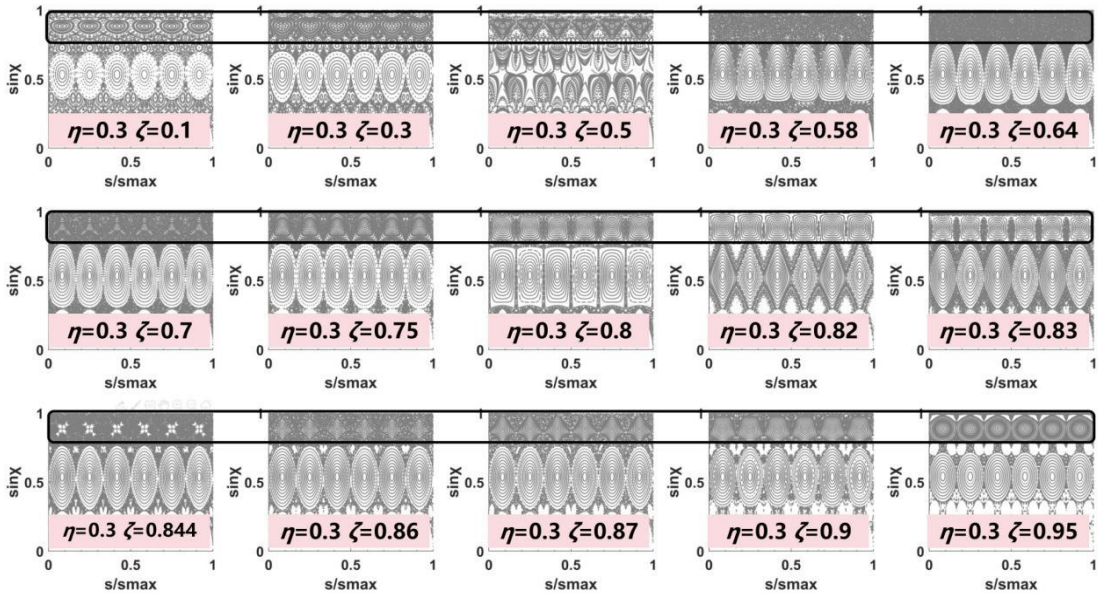

Fig. S3. The full PSOSs of hexagonal cavities when keeping $\eta = 0.3$ and varying $\zeta$ as 0.1, 0.3, 0.5, 0.58, 0.64, 0.7, 0.75, 0.8, 0.82, 0.83, 0.844, 0.86, 0.87, 0.9 and 0.95. The 6-period motion is enclosed in black rectangular boxes in the PSOSs.

*Contact author: felix831204@xjtu.edu.cn

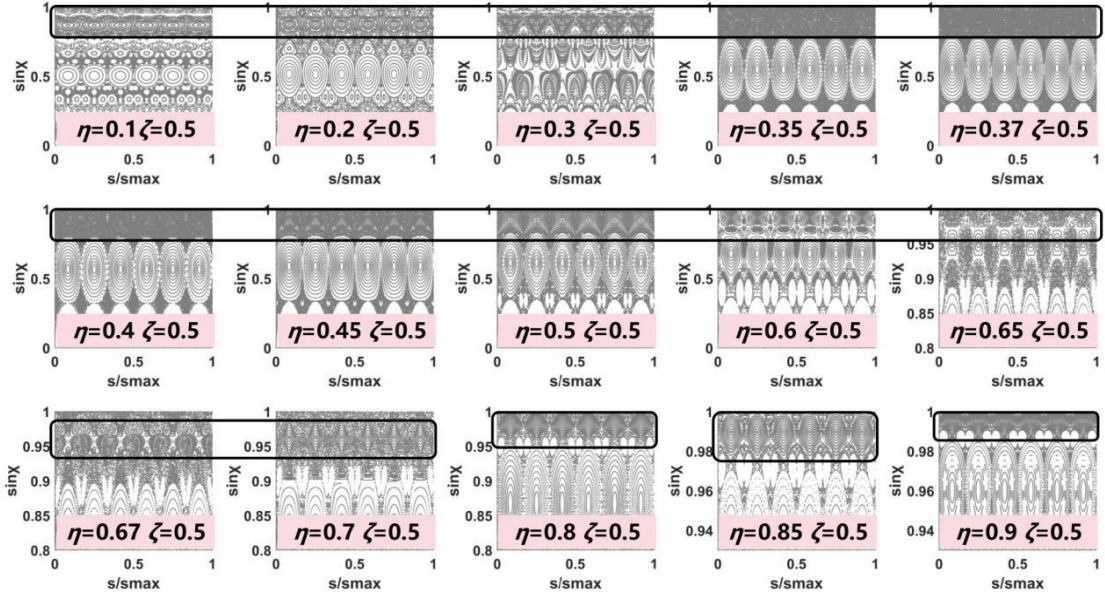

Fig. S4. The full PSOSs of hexagonal cavities when keeping $\zeta = 0.5$ and varying $\eta$ as 0.1, 0.2, 0.3, 0.35, 0.37, 0.4, 0.45, 0.5, 0.6, 0.65, 0.67, 0.7, 0.8, 0.85 and 0.9 The 6-period motion is enclosed in black rectangular boxes in the PSOSs.

2. **The phase diagram of fixed points for polygons with various numbers of sides**.

We present the phase diagram of fixed points for triangular, quadrilateral, pentagonal, hexagonal, and octagonal cavities. It is demonstrated that the blue dashed line, representing the 8-branch hyperbolic fixed point, intersects with the lower right corner of the phase diagram for quadrilateral cavities, while the red dashed line, representing the 6-branch hyperbolic fixed point, intersects with the lower right corner of the phase diagram for hexagonal cavities.

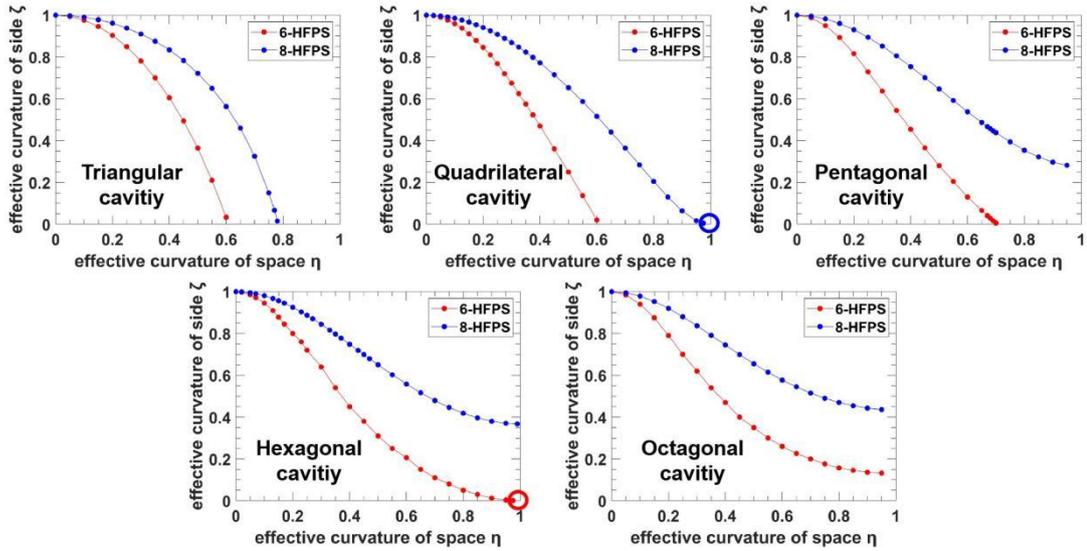

Fig. S5. Phase diagrams for triangular, quadrilateral, pentagonal, hexagonal and octagonal cavities. The red dashed lines represent 6-HFPs and the blue dashed lines represent 8-HFPs. (The phase diagram curves of the quadrilateral microcavity and hexagonal microcavity intersecting

*Contact author: felix831204@xjtu.edu.cn

with the phase point ($\eta=1$, $\zeta=0$) are marked by blue and red circles, respectively.)

3. **The mathematics of transformation optics.**

To preserve the invariance of the geodesic equation and the wave equation, it is necessary to perform a transformation from 3D cavities with a uniform refractive index to 2D cavities with a gradient refractive index, such that the conditions of conformal mapping are satisfied. Specifically, this requires that

$$n_{2D}^2(r) \cdot ds_{2D}^2 = n_{3D}^2 \cdot ds_{3D}^2 \tag{S1}$$

Where $ds_{3D}^2$ and $ds_{2D}^2$ are the line elements of 3D and 2D microcavities respectively. And for 3D hemispherical polygonal microcavity, we can use spherical coordinates to present

$$ds_{3D}^2 = R_0^2(d\theta^2 + \sin^2\theta d\varphi^2) \tag{S2}$$

Where $R_0$ is spherical radius, $\varphi$ is rotation angle and $\theta$ is elevation of the sphere. For 2D polygonal microcavity, we can use polar coordinates to present

$$ds_{2D}^2 = dr^2 + r^2 d\phi^2 \tag{S3}$$

Where $r$ is polar diameter and $\phi$ is polar angle. Thus, by inserting Eq. (S2) and Eq. (S3) to Eq. (S1), we can get

$$\tilde{n}^2(dr^2 + r^2 d\phi^2) = R_0^2(d\theta^2 + \sin^2\theta d\varphi^2) \tag{S4}$$

Where $\tilde{n} = n_{2D}/n_{3D}$, and it is obvious that $\phi$ in 2D cavity and $\varphi$ in 3D cavity represent the same physical quantity, thus we can know that

$$\tilde{n} dr = R_0 d\theta \tag{S5}$$

$$\tilde{n} r = R_0 \sin\theta \tag{S6}$$

By combining Eq. (S5) and Eq. (S6), we get the relation between polar diameter $r$ in 2D and elevation of the hemispherical cavity $\theta$

$$r = C \cdot \tan(\frac{\theta}{2}) \tag{S7}$$

By inserting Eq. (S7) into Eq. (S6), we can get

$$n_{2D} = n_{3D} \cdot \frac{2 \cdot R_0/C}{(r/C)^2 + 1} \tag{S8}$$

In this work, we require that the size of the cavities remains unchanged under the geometric transformation, therefore, we keep $R_e = R_0 \cdot \sin(\eta \cdot \pi/2)$ as a constant. Thus, when $\theta_{max} = \eta \cdot \pi/2$, let $r_e = R_e$, then we can know

$$C = R_e / \tan(\eta \cdot \pi/4) \tag{S9}$$

*Contact author: felix831204@xjtu.edu.cn